\documentclass{article}%
      \usepackage{gsirep}%
      \usepackage{graphicx}%
      \title{Evidence for $\psi'$ Regeneration in Heavy Ion Collisions}
      \author[1,2]{Andriy Kostyuk}
      \author[1]{Horst St\"ocker}
      \affil[1]{{\it Institut f\"ur Theoretische Physik and Frankfurt 
			Institute for Advanced Studies, J.W. Goethe-Universit\"at,  
			Frankfurt am Main, Germany}}
       \affil[2]{{\it Bogolyubov Institute for Theoretical Physics, Kyiv,
       Ukraine}}

      \date{28.02.2004}
      \begin{document}
      \maketitle

The study of hidden charm production is an important part of the
heavy ion program.  The standard approach to this problem
\cite{MS} assumes that $c\bar{c}$ bound states  are created {\it
only} at the initial stage of the reaction and then partially
destroyed at later stages due to interactions with the medium
\cite{Blaizot,Capella,Spieles}.

The idea of the statistical $J/\psi$ production \cite{GG} triggered
the development of an alternative approach \cite{BMS,We}  --- the
statistical coalescence model (SCM): charmonia (as well as open
charm hadrons) are formed at hadronization  due to coalescence of
charm quarks and antiquarks.


The standard approach (two its versions --- the threshold
suppression model \cite{Blaizot} and comover model \cite{Capella,Spieles})
demonstrates reasonable agreement with the data on $J/\psi$
production in Pb+Pb collisions at SPS. Better agreement, but
only for (semi)central collisions $N_p > 100$ ($N_p$ is the number
participant nucleons) can be obtained within the SCM \cite{charm7}.

\begin{figure}
\includegraphics[width=9 cm]{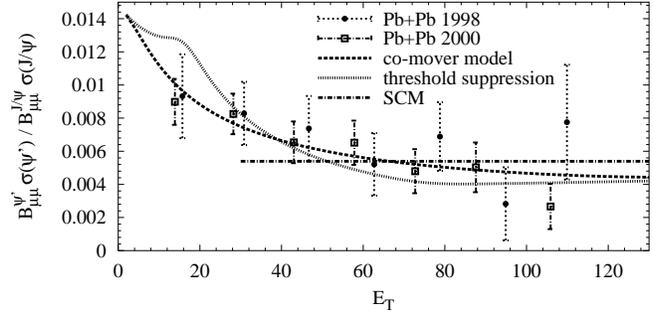}
\vspace*{-0.8cm}
      \caption{The $\psi'$ to $J/\psi$ ratio in Pb+Pb collisions at SPS.}
\label{fig1} \vspace*{-0.2cm}
\end{figure}

It is interesting to check, which of the two scenarios is better
suited for excited charmonium states. Recently the N50 collaboration
presented new data on $\psi'$ production in Pb+Pb collisions
\cite{NA50Pb}.  It appears that the both versions of the standard
approach as well as SCM are able to fit the new $\psi'$ data (see
Fig. \ref{fig1}). But the observed $\psi'$ suppression seems to be too
weak to give reasonable values of the fit parameters of the
standard scenario.

The free parameter of the threshold suppression model is the
threshold density of the participant nucleons in the transverse
plane at which the charmonium species under consideration is
suppressed. The $J/\psi$ data can be reasonably fitted with
$n_{J/\psi}=3.77
\begin{array}{l}
{ \scriptstyle + 0.09} \\[-1.5ex]
{ \scriptstyle - 0.10}
\end{array} $ fm$^{-2}$
for primary $J/\psi$'s and
$n_{\chi}=1.95
\begin{array}{l}
{ \scriptstyle + 0.35} \\[-1.5ex]
{ \scriptstyle - 0.45}
\end{array}
$ fm$^{-2}$ for excited charmonia (mostly $\chi$-states)
contributing up to 40\% to the total $J/\psi$ yield in p+p collisions. 
Because the 
$\psi'$ is much weaker bound than any of the $\chi$ states, one would
expect that its suppression begins at substantially lower density.
It appears, however, that the $\psi'$ data suggest approximately
the same (or even larger) threshold for the $\psi'$ suppression as
for $\chi$:
$n_{\psi'}=2.17
\begin{array}{l}
{ \scriptstyle + 0.33} \\[-1.5ex]
{ \scriptstyle - 0.43}
\end{array} $ fm$^{-2}$.

The comover model does not alow to extract the suppression
parameters for primary $J/\psi$ and $\chi$ separately. The free
parameter of the model is the {\it effective} suppression  cross
section averaged over all charmonium states that contribute to the
production of the species under consideration. There is no obvious
contradiction between the fit results for $J/\psi$ and $\psi'$:
$\sigma^{co}_{J/\psi}=1.01 \pm 0.05$~mb and
$\sigma^{co}_{\psi'}=2.84
\begin{array}{l}
{ \scriptstyle + 1.56} \\[-1.5ex]
{ \scriptstyle - 0.79}
\end{array}$~mb, i.e. the comover dissociation cross section for
$\psi'$ is by a factor of about $3$ larger than that for $J/\psi$.
This ratio of the cross sections can be obtained, if one assumes
that the matrix element of the reaction is approximately the same
for $J/\psi$ and $\psi'$ \cite{brat} and the suppression is dominated by
exothermic (e.g. $J/\psi + \rho \rightarrow D+D$) reactions.
Still, model calculations of the
matrix element predict much larger 
dissociation cross section for $\psi'$ than for $J/\psi$
\cite{models}. If that is indeed so, then the only possible
explanation of the observed $\psi'$ yield is regeneration of
$\psi'$ at late stages of the reaction. As can be seen from Fig. \ref{fig1}
the new data (as well as the old ones \cite{BMS,Shur}) are
consistent with the assumption that the $\psi'$ to $J/\psi$ ratio
is nearly constant at $N_p > 100-150$. It is equal to the thermal value,
corresponding to freeze-out temperatures $T=150-170$ MeV, which
agrees with the SCM. This suggest that excited  charmonium states are
likely to be produced at the final rather than at the initial stage of
heavy ion reactions.

Measurement of the charmonium production in heavy ion collisions
at lower energies, e.g. at the new GSI accelerator facility, would help
to disentangle different charmonium production mechanisms.

We acknowlege the finantial support of GSI, DFG and BMBF.

\end{document}